\documentclass[aps,twocolumn,showpacs,preprintnumbers,floats]{revtex4}\def\@cite#1#2{\textsuperscript{[{#1\if@tempswa , #2\fi}]}}
\usepackage{mathrsfs}

\usepackage{longtable,lscape}
\usepackage{savesym}
\usepackage{txfonts}
\savesymbol{iint}
\savesymbol{iiint}
\savesymbol{iiiint}
\savesymbol{idotsint}
\usepackage{amssymb}
\usepackage{amsmath}
\restoresymbol{TXF}{iint}
\restoresymbol{TXF}{iiint}
\restoresymbol{TXF}{iiiint}
\restoresymbol{TXF}{idotsint}
\usepackage{indentfirst}
\usepackage{graphicx,booktabs}
\usepackage{braket}
\usepackage{multirow}
\usepackage{color}
\usepackage{subfigure}
\usepackage{float}
\usepackage{bm}
\usepackage{epsfig}
\usepackage[colorlinks, citecolor=blue,anchorcolor=red,menucolor=red, linkcolor=red,filecolor=red,urlcolor=blue,frenchlinks=red]{hyperref}
\newcommand{\vsig}{\mbox{\boldmath$\sigma$\unboldmath}}

\begin{document}
	
	\title{Strong decays of the low-lying $1P$- and $1D$-wave $\Sigma_c$ baryons}
	\author{Yu-Hui Zhou$^{1}$, Wen-Jia Wang$^{1}$, Li-Ye Xiao$^{1}$~\footnote {E-mail: lyxiao@ustb.edu.cn}, Xian-Hui Zhong$^{2,3}$~\footnote {E-mail: zhongxh@hunnu.edu.cn}}
\affiliation{ 1)Institute of Theoretic
al Physics, University of Science and Technology Beijing,
Beijing 100083, China}
\affiliation{ 2) Department of
Physics, Hunan Normal University, and Key Laboratory of
Low-Dimensional Quantum Structures and Quantum Control of Ministry
of Education, Changsha 410081, China }
\affiliation{ 3) Synergetic
Innovation Center for Quantum Effects and Applications (SICQEA),
Hunan Normal University, Changsha 410081, China}
	
\begin{abstract}
In this work, we systematically study the OZI-allowed two-body strong decay properties of $1P$- and $1D$-wave $\Sigma_c$ baryons within the $j $-$j$ coupling scheme in the framework of the quark pair creation model. For a comparison, we also give the predictions of the chiral quark model.
Some model dependencies can be found in the predictions of two models. The calculations indicate that: ($\rm{\romannumeral1}$) The $1P$-wave $\lambda$-mode $\Sigma_c$ states most likely to be relatively narrow states with a width of $\Gamma<80$ MeV. Their main
decay channels are $\Lambda_c\pi$, or $\Sigma_c\pi$, or $\Sigma_c^*\pi$. The $1P$-wave $\rho$-mode states most might be broad states with a width of $\Gamma\sim 100-200$ MeV. They dominantly decay into $\Sigma_c\pi$ and $\Sigma_c^*\pi$ channels.
Some evidences of these $1P$-wave states are most likely to be observed in the $\Lambda_c\pi$ and $\Lambda_c\pi\pi$
invariant mass spectra around the energy range of $2.75-2.95$ GeV.
($\rm{\romannumeral2}$) The $1D$-wave $\lambda$-mode $\Sigma_c$ excitations may be moderate states with a width of about dozens of MeV.
The $1D$-wave $\lambda$-mode states mainly decay into the $1P$-wave charmed baryon via the pionic decay processes.
Meanwhile, several $1D$-wave $\lambda$-mode states have significant decay rates into $DN$ or $D^*N$. Hence, the $DN$ and $D^*N$ are likely to be interesting channels for experimental exploration. ($\rm{\romannumeral3}$) Furthermore,
the two $1D$-wave $\rho$-mode excitations $\Sigma_c|J^P=5/2^+,3\rangle_{\rho\rho}$ and $|J^P=7/2^+,3\rangle_{\rho\rho}$
are most likely to be fairly narrow state with a width of dozens of MeV, and they mainly decay into
$\Lambda_c\pi$. Some evidences of them might be observed in the $\Lambda_c\pi$ invariant mass spectra around the energy range of $3.1-3.2$ GeV.
\end{abstract}
	\maketitle

\section{Introduction}

In the past five years, a great progress has been achieved on the observation of singly heavy baryons in experiment. For the baryons containing single bottom quark, there are almost twelve new structures recently observed by LHCb and CMS. In 2018, $\Sigma_b(6097)^{\pm}$~\cite{LHCb:2018haf} was reported by LHCb Collaboration in the $\Lambda_b^0\pi^{\pm}$ systems. At the same year, the LHCb Collaboration again found another resonance, $\Xi_b(6227)^-$, both
in the $\Lambda_b^0K^-$ and $\Xi_b^0\pi^-$ channels~\cite{LHCb:2018vuc}. In 2019, two almost degenerate narrow states $\Lambda_b(6146)^0$ and $\Lambda_b(6152)^0$ were observed by LHCb in the $\Lambda_b\pi^+\pi^-$ invariant mass distribution~\cite{LHCb:2019soc}. Especially in 2020,
the LHCb Collaboration reported four narrow peaks, $\Omega_b(6316)^-$, $\Omega_b(6330)^-$, $\Omega_b(6340)^-$ and $\Omega_b(6350)^-$ in the $\Xi_b^0K^-$ channel~\cite{LHCb:2020tqd}. Moreover, in 2021, the CMS Collaboration observed a narrow resonance $\Xi_b(6100)^-$ in the $\Xi_b^-\pi^+\pi^-$ invariant mass spectrum~\cite{CMS:2021rvl}. Very recently, the LHCb Collaboration reported two new excited $\Xi_b^0$ states $\Xi_b(6327)^0$ and $\Xi_b(6333)^0$ in the $\Lambda_b^0K^-\pi^+$ mass spectrum~\cite{LHCb:2021ssn}.
On other hand, for the singly charmed baryons, the progress in experiment is also surprising. In 2017, the LHCb Collaboration observed five new narrow $\Omega_c^0$ states~\cite{LHCb:2017uwr}: $\Omega_c(3000)^0$, $\Omega_c(3050)^0$, $\Omega_c(3066)^0$, $\Omega_c(3090)^0$ and $\Omega_c(3119)^0$, of which all decay strongly to the final state $\Xi_c^+K^-$. The first four of them were confirmed by  Belle experiments subsequently~\cite{Belle:2017ext}. In addition, at the same year the LHCb Collaboration investigated a near-threshold enhancement in the $D^0p$ amplitude and found a new $\Lambda_c^+$ resonance, denoted the $\Lambda_c(2860)^+$~\cite{LHCb:2017jym}. This new state has been predicted as well by some references before~\cite{Chen:2016iyi,Chen:2016phw,Lu:2016ctt}. Later, the LHCb
Collaboration reported three new $\Xi_c^0$ states, $\Xi_c(2923)^0$, $\Xi_c(2939)^0$ and $\Xi_c(2965)^0$ in $\Lambda_c^+K^-$ decay mode~\cite{LHCb:2020iby} with no-determined spin-parity. Not long ago the Belle collaboration found possibly an excited $\Lambda_c^+$ state, $\Lambda_c(2910)^+$, via studying $\bar{B}^0\rightarrow\Sigma_c(2455)^{0,++}\pi^{\pm}\bar{p}$ decays~\cite{Belle:2022hnm}. Recently, the LHCb collaboration observed two new excited states, $\Omega_c(3185)^0$ and $\Omega_c(3327)^0$~\cite{LHCb:2023rtu}, in the $\Xi_c^+K^-$ invariant-mass spectrum.

To decode the inner structures of those newly observed singly heavy baryons, there exist many theoretical calculations of the mass spectra and strong decays with various models in the literature~\cite{Wang:2017hej,Cheng:2017ove,Chen:2018orb,Chen:2017gnu,Bijker:2020tns,Xiao:2020gjo,
Wang:2020gkn,Wang:2018fjm,Lu:2020ivo,Wang:2017vnc,Padmanath:2017lng,Karliner:2017kfm,Wang:2017zjw,Yang:2019cvw,Cui:2019dzj,
Yang:2020zrh,Aliev:2018vye,Chen:2018vuc,Yang:2018lzg,Jia:2019bkr,He:2021xrh,Wang:2020pri,Mutuk:2020rzm,Liang:2020hbo,
Agaev:2017lip,Agaev:2020fut,Huang:2017dwn,Kishore:2019fzb,Huang:2018bed,Wang:2021cku,Hu:2020zwc}. The most popular explanation is interpreting those newly observed structures into traditional hadronic excited states, such as $1P$-wave ~\cite{Chen:2018orb,Chen:2017gnu,Bijker:2020tns,Xiao:2020gjo,Wang:2020gkn,Wang:2018fjm,Wang:2017vnc,Padmanath:2017lng,
Karliner:2017kfm,Yang:2019cvw,Cui:2019dzj,Yang:2020zrh,Aliev:2018vye,Chen:2018vuc,Yang:2018lzg,
Jia:2019bkr,He:2021xrh,Wang:2020pri,Mutuk:2020rzm,Liang:2020hbo,Agaev:2017lip}, $2P$-wave~\cite{Cheng:2017ove}, $1D$-wave~\cite{Cheng:2017ove} or $2S$-wave~\cite{Wang:2017hej,Cheng:2017ove,Lu:2020ivo,Agaev:2020fut,Huang:2017dwn} excitations. Meanwhile, there are other unconventional interpretations, such as molecular states~\cite{Huang:2017dwn,Kishore:2019fzb,Liu:2018bkx,Yu:2018yxl,Wang:2020vwl} or compact pentaquark states~\cite{Wang:2021cku,Hu:2020zwc}.

The $\Sigma_c$ baryon spectrum is one of important members of the singly heavy baryons. While, there are very few signals of $\Sigma_c$
baryons from experiments. Except for the well-established ground states, $\Sigma_c(2455)$ and $\Sigma_c(2520)$~\cite{ParticleDataGroup:2020ssz}, there is just one controversial candidate of excited $\Sigma_c$ baryon, denoted as $\Sigma_c(2800)$~\cite{Belle:2004zjl}, of which the spin-parity are not determined. Fortunately, the upgrades of experimental technology of the LHC experiments have demonstrated
a powerful capability in discovering the heavy baryons. Therefore, the missing $1P$- and $1D$-wave $\Sigma_c$ excitations are likely to be discovered by forthcoming experiments. The mass spectrum of the $\Sigma_c$ baryons were studied extensively in the literature~\cite{Capstick:1986ter,Yu:2022ymb,Ebert:2011kk,Ebert:2007nw,Migura:2006ep,Yoshida:2015tia,Shah:2016nxi,Jia:2019bkr,Chen:2016iyi,
 Bahtiyar:2020uuj,Perez-Rubio:2015zqb,Chen:2015kpa}. It should be mentioned that for an excited singly heavy baryon,
there are two kinds of excitations: $\lambda$-mode and $\rho$-mode excitations.
The $\lambda$-mode excitation appears between the light diquark and the heavy quark,
while the $\rho$-mode excitation occurs within the light diquark. According to the quark model
predictions~\cite{Bijker:2020tns,Yoshida:2015tia,Capstick:1986ter,Chen:2021eyk} the mass of the $\lambda$-mode
resonances is about $70-150$ MeV lower than that of $\rho$-mode resonances. We collect the mass predictions of the
$1P$- and $1D$-wave $\Sigma_c$ resonances in Table~\ref{table1}. Except mass spectrum, the strong decay properties of
the $\Sigma_c$ baryons were also studied with various theoretical methods and models, such as the chiral
quark model (ChQM)~\cite{Wang:2021bmz,Yao:2018jmc}, the quark pair creation (QPC) model~\cite{Garcia-Tecocoatzi:2022zrf},
lattice QCD~\cite{Can:2016ksz}, QCD sum rules~\cite{Azizi:2008ui}, Eichten-Hill-Quigg (EHQ) formula~\cite{Chen:2016iyi},
heavy hadron chiral perturbation theory~\cite{Cheng:2015naa} and non-relativistic quark model~\cite{Nagahiro:2016nsx,Albertus:2005zy}, and so on.

\begin{table*}
	\caption{Predicted mass of $\Sigma_c$ baryons($1P$-wave  and $1D$-wave)  in various quark models and possible decay channels within the QPC model.}	
	\label{table1}
	\begin{tabular}{cccccccccccccc}
		\toprule[1.2pt]
\multirow{1}{*}{Notation}& \multicolumn{6}{c}{$\rm{Quantum}$ $\rm{Number}$}& \multicolumn{6}{c}{$\rm{Mass}$}  & \multicolumn{1}{c}{$\rm{Decay}$ $\rm{channel}$}  \\
				\hline
		$\Sigma_c$${\ket{J^P,j}}_{\lambda(\rho)}$&$l_{\lambda}$&$l_{\rho}$&$L$&$s_{\rho}$&$j$&$J^P$&GIM~\cite{Yu:2022ymb}&RQM~\cite{Ebert:2011kk}& RQM~\cite{Ebert:2007nw}& NQM~\cite{Chen:2016iyi}&NQM~\cite{Yoshida:2015tia}&QCD~\cite{Chen:2015kpa}& \\
		
		\hline
		$\Sigma_c$${\ket{J^P=\frac{1}{2}^-,0}}_\lambda$&1&0&1&1&0&$\frac{1}{2}^{-}$
&$2823$ &$2799 $&$2805$ &$2702$&2802 &$2820$&$\Sigma^{(*)}_c\pi$,$\Lambda_c\pi$,$DN$ \\
		$\Sigma_c$${\ket{J^P=\frac{1}{2}^-,1}}_\lambda$  &1&0&1&1&1&$\frac{1}{2}^{-}$	&$2809$ &$2713 $&$2795$ &$2765$ &2826&$2790$  \\	

	$\Sigma_c$${\ket{J^P=\frac{3}{2}^-,1}}_\lambda$&1&0&1&1&1&$\frac{3}{2}^{-}$
		&$2829$ &$2798 $&$2799$ &$2785$&2807 &$2820$\\
		$\Sigma_c$${\ket{J^P=\frac{3}{2}^-,2}}_\lambda$&1&0&1&1&2&$\frac{3}{2}^{-}$  &$2802$ &$2773 $&$2761$ &$2798$ &2837&$2800$\\
	
	$\Sigma_c$${\ket{J^P=\frac{5}{2}^-,2}}_\lambda$	&1&0&1&1&2&$\frac{5}{2}^{-}$
		&$2835$ &$2789 $&$2790$ &$2790$ &2839&$2890$   \\
		\hline
		$\Sigma_c$${\ket{J^P=\frac{1}{2}^-,1}}_\rho$	&0&1&1&0&1&$\frac{1}{2}^{-}$&&&&&2909& &$\Sigma^{(*)}_c\pi$  \\
		$\Sigma_c$${\ket{J^P=\frac{3}{2}^-,1}}_\rho$	&0&1&1&0&1&$\frac{3}{2}^{-}$
		& && & &2910&  \\
		\hline
	$\Sigma_c$${\ket{J^P=\frac{1}{2}^+,1}}_{\lambda\lambda}$&2&0&2&1&1&$\frac{1}{2}^{+}$
	&$3073$ &3041&3014&2949&3103&&$\Sigma^{(*)}_c\pi$,$\Lambda_c\pi$,$D^{(*)}N$,$\Xi_cK$,$\Xi^{'}_cK$,$|\Lambda_c(\Sigma_c)P_{\lambda}\rangle\pi$ \\

	$\Sigma_c$${\ket{J^P=\frac{3}{2}^+,1}}_{\lambda\lambda}$&2&0&2&1&1&$\frac{3}{2}^{+}$  	
	&$3084$ &3043&3005&2952&3065 \\
	$\Sigma_c$${\ket{J^P=\frac{3}{2}^+,2}}_{\lambda\lambda}$&2&0&2&1&2&$\frac{3}{2}^{+}$  	
&$3073$ &3040&3010&2964&3094 \\
$\Sigma_c$${\ket{J^P=\frac{5}{2}^+,2}}_{\lambda\lambda}$	&2&0&2&1&2&$\frac{5}{2}^{+}$  	
&$3085$ &3038&3001&2942&3099 \\
$\Sigma_c$${\ket{J^P=\frac{5}{2}^+,3}}_{\lambda\lambda}$	&2&0&2&1&3&$\frac{5}{2}^{+}$  	
&$3072$ &3023&2960&2962&3114 \\
$\Sigma_c$${\ket{J^P=\frac{7}{2}^+,3}}_{\lambda\lambda}$	&2&0&2&1&3&$\frac{7}{2}^{+}$  	
&$3086$ &3013&3015&2943&  \\
	\hline
$\Sigma_c$${\ket{J^P=\frac{1}{2}^+,1}}_{\rho\rho}$&0&2&2&1&1&$\frac{1}{2}^{+}$ &&&&&&&$\Sigma^{(*)}_c\pi$,$\Lambda_c\pi$,$\Xi_cK$,$\Xi^{'}_cK$,$|\Lambda_c(\Sigma_c)P_{\lambda}\rangle\pi$ \\
$\Sigma_c$${\ket{J^P=\frac{3}{2}^+,1}}_{\rho\rho}$&0&2&2&1&1&$\frac{3}{2}^{+}$ &&&&&&&\\
$\Sigma_c$${\ket{J^P=\frac{3}{2}^+,2}}_{\rho\rho}$&0&2&2&1&2&$\frac{3}{2}^{+}$ &&&&&&&\\
$\Sigma_c$${\ket{J^P=\frac{5}{2}^+,2}}_{\rho\rho}$&0&2&2&1&2&$\frac{5}{2}^{+}$ &&&&&&&\\
$\Sigma_c$${\ket{J^P=\frac{5}{2}^+,3}}_{\rho\rho}$&0&2&2&1&3&$\frac{5}{2}^{+}$ &&&&&&&\\
$\Sigma_c$${\ket{J^P=\frac{7}{2}^+,3}}_{\rho\rho}$&0&2&2&1&3&$\frac{7}{2}^{+}$ &&&&&&&\\
		\bottomrule[1.2pt]
	\end{tabular}
\end{table*}

For the low-lying $1P$-wave and $1D$-wave $\Sigma_c$ baryons, their masses allow them decaying
into the $DN$ and $D^*N$ ($N$ denoting a nucleon) channels.
However, only a few discussions exist in literature~\cite{Yu:2022ymb}. Meanwhile, the study on strong decay properties of
$\rho$-mode excitations is rare. Based on those actuality, in the present work we carry out
a systematic analysis of the $1P$- and $1D$-wave $\Sigma_c$ states for both $\rho$- and
$\lambda$-mode excitations with the QPC model.
For a comparison, we also give the predictions within the the chiral quark model.
The quark model classification and Okubo-Zweig-Iizuka (OZI)
allowed two-body decay channels are summarized in Table~\ref{table1} as well.
It should be mentioned that the spatial wave function of hadrons is adopted the harmonic oscillator form in this work,
the $D^{(*)}N$ decay channels for the $\rho$-mode excitations are forbidden. Hence, for the $\rho$-mode excitations,
we only focus on the decay channels containing light $\pi$ and $K$ mesons.

This paper is structured as follows. In Sec. II, we briefly
introduce the QPC model, chiral quark model and the relationship of states in different coupling schemes.
Then we present our numerical results and discussions in Sec. III. A summary is given in Sec. IV.

\section{Theoretical framework}

\subsection{QPC model}\label{model}

The QPC model, also famous as the $^3P_0$ model, was first proposed by Micu \cite{Micu:1968mk}, Car-litz and Kislinger \cite{R1970Regge}, and further developed by the Orsay group \cite{LeYaouanc:1972vsx,LeYaouanc:1977fsz,LeYaovanc1988Hadron}.
The main idea of this model is that strong decay takes place via the creation of quark-antiquark pair from the vacuum with quantum number $0^{++}$ .
This model has been extensively applied to study the OZI-allowed strong transitions of hadron systems \cite{Chen:2007xf,Zhao:2016qmh,Chen:2017aqm,Limphirat:2010zz,He:2021xrh,Lu:2018utx}.

\begin{figure}[h]
	\begin{center}
		\subfigure[]{\begin{minipage}{0.32\linewidth}
				\centering
				\includegraphics[width=1in,height=1in]{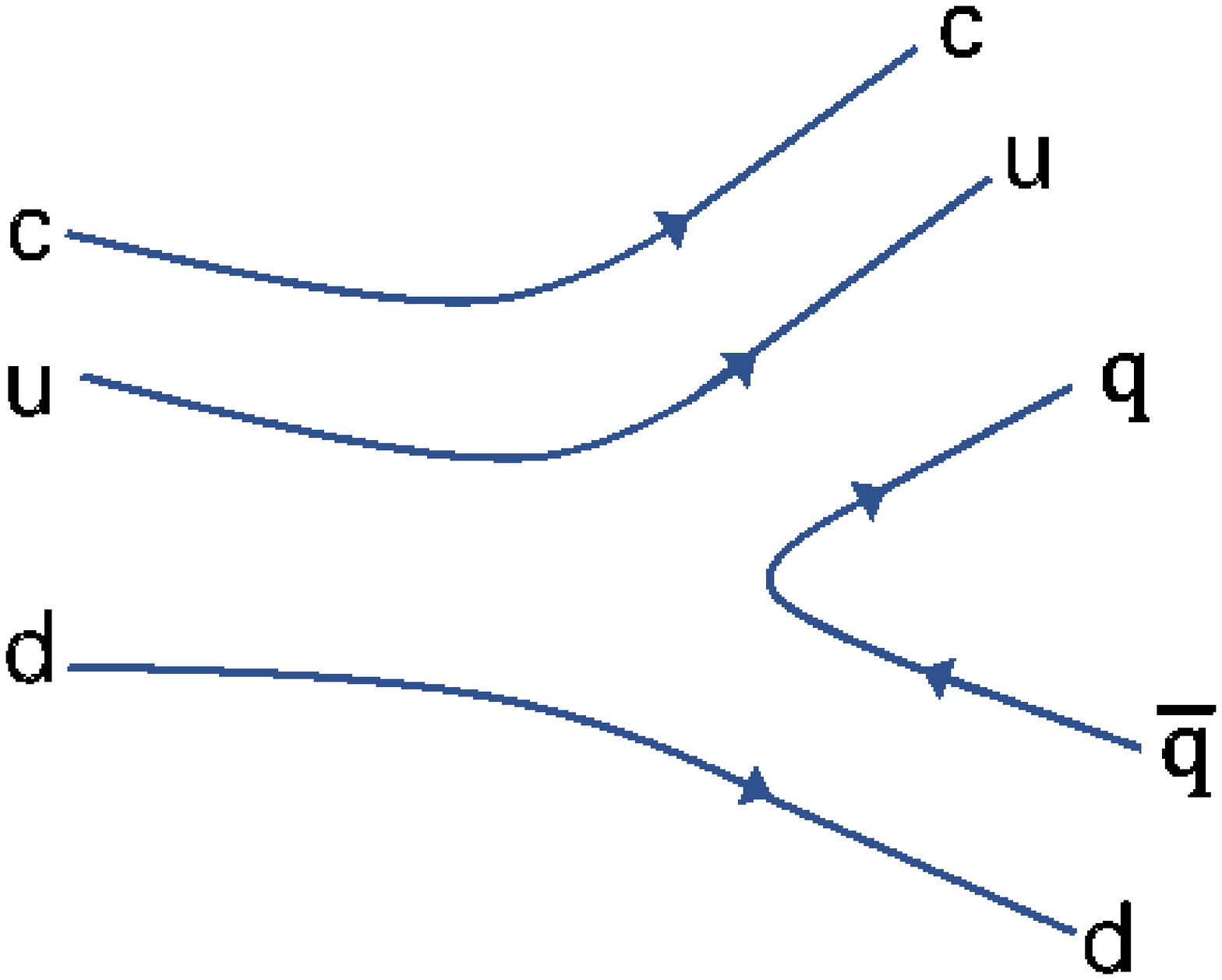}
		\end{minipage}}
		\subfigure[]{\begin{minipage}{0.32\linewidth}
				\centering
				\includegraphics[width=1in,height=1in]{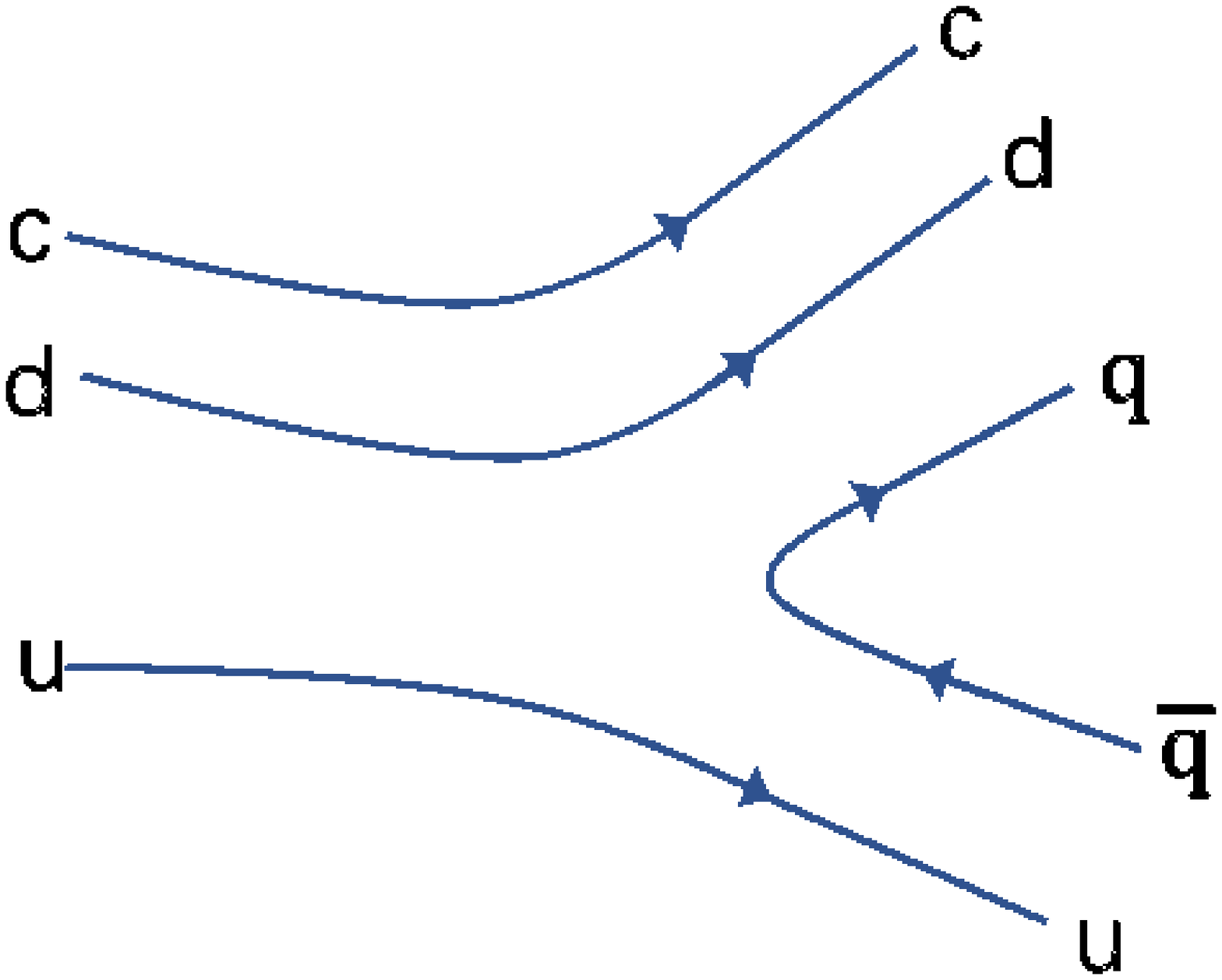}
		\end{minipage}}
		\subfigure[]{\begin{minipage}{0.32\linewidth}
				\centering
				\includegraphics[width=1in,height=1in]{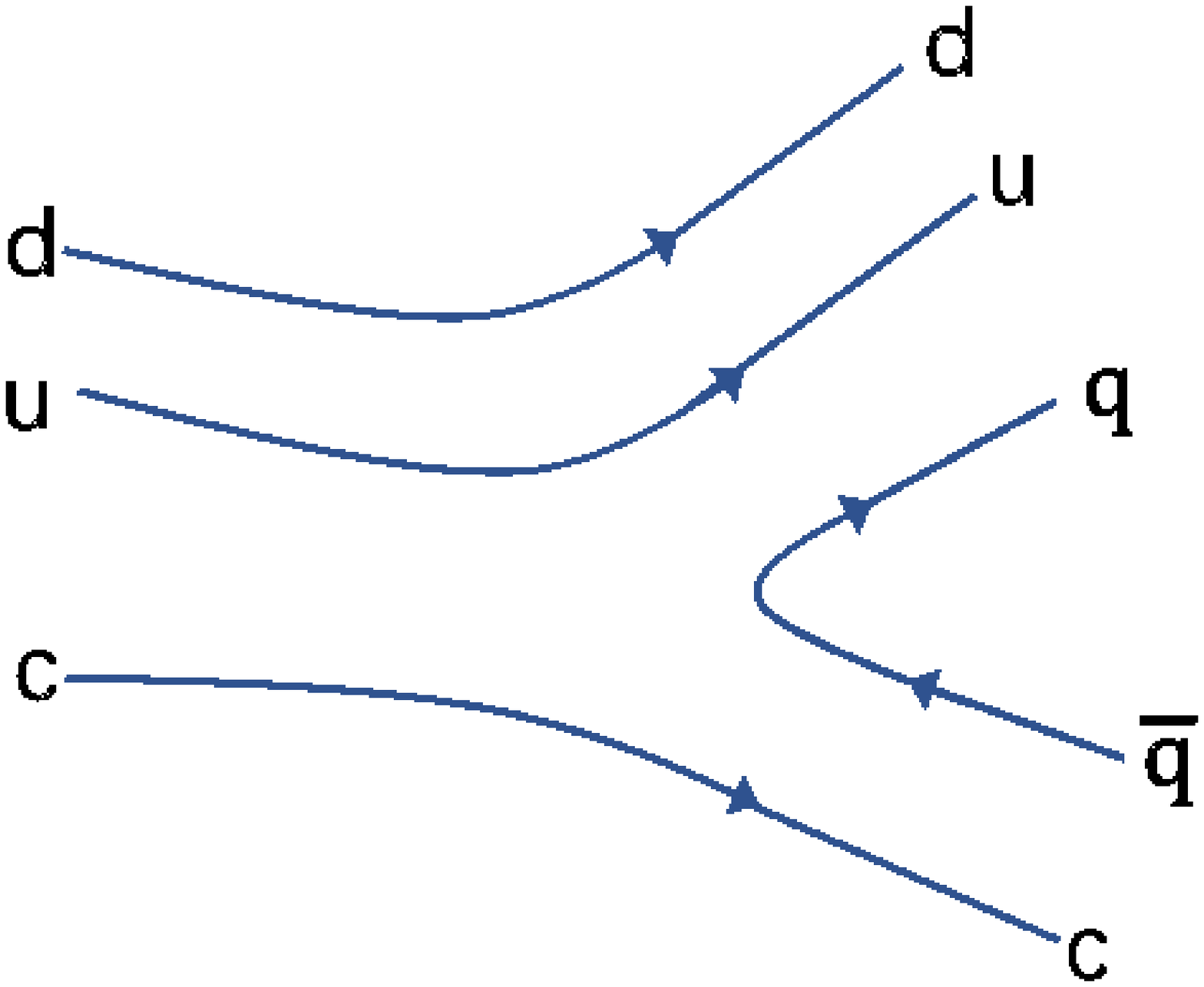}
		\end{minipage}}
		\centering
		\caption{Decay process of an excited $\Sigma_c$ baryon in the QPC model.}\label{FIG1}
	\end{center}
\end{figure}

For a baryon decay process, one quark of the initial baryon state $|A\rangle$ regroups with the
created antiquark to form a meson $|C\rangle$, and the other two quarks regroup with the created quark to
form a daughter baryon $|B\rangle$.
Considering the $\Sigma_c^+$ baryons being composed of two light quarks ($u$ and $d$) and a heavy quark $c$, thus,
according to the quark rearrangement process, any of the three quarks in the initial baryon can go into the final meson. Hence, there are three possible decay ways, as shown in Fig~\ref{FIG1}.

In the QPC model, the transition operator for the OZI-allowed
two-body decay process $A\to BC$ is given by
\begin{equation}
	\begin{aligned}
	T=&-3\gamma\sum\limits_{m}\langle1m;1-m|00\rangle\int d^3\textbf{p}_4d^3\textbf{p}_5\delta^3(\textbf{p}_4+\textbf{p}_5)\\ &\times\mathcal{Y}^m_1(\dfrac{\textbf{p}_4-\textbf{p}_5}{2})\chi^{45}_{1-m}\phi^{45}_0\omega^{45}_0a^\mathcal{y}_{4i}(\textbf{p}_4)b^\mathcal{y}_{5j}(\textbf{p}_5),\\
\end{aligned}
\end{equation}
where $\gamma$ is a dimensionless parameter and accounts for the vacuum pair-production strength.
$\textbf{p}_4$ and  $\textbf{p}_5$ are the three-vector momenta of the created quark  pair.
$\omega^{45}_0$=$\delta_{ij}$,
$\phi^{45}_0$=(u$\rm{\overline{u}}$+d$\rm{\overline{d}}$+s$\rm{\overline{s}}$)/$\sqrt{3}$ and $\chi^{45}_{1-m}$ denote the color singlet, flavor function and spin triplet of the quark pair, respectively.
The solid harmonic polynomial $\mathcal{Y}^m_1$=$|\textbf{p}|$$\rm{\boldsymbol{Y}}^m_1$($\theta_p\phi_p$) reflects the momentum-space distribution .
The creation operator $a^\mathcal{y}_{4i}d^\mathcal{y}_{5j}$ denotes the quark pair-creation in the vacuum.

According to the definition of the mock state in the constituent quark model~\cite{Hayne:1981zy}, the wave
functions of the baryon states $|A\rangle$ and $|B\rangle$ and meson state $|C\rangle$ can be easily
obtained. For example for the initial baryon state $|A\rangle$, the wave function is given by
\begin{equation}
	\begin{aligned}
		&|A(N^{2S_A+1}_AL_AJ_AM_{J_A})(\textbf{P}_A)\rangle\\
		=&\sqrt{2E_A}\varphi^{123}_A\omega^{123}_A\sum\limits_{M_{L_A},M_{S_A}}\langle L_AM_{L_A};S_AM_{S_A}|J_A,M_{J_A}\rangle\\
		&\times\int d^3\textbf{p}_1d^3\textbf{p}_2d^3\textbf{p}_3\delta^3(\textbf{p}_1+\textbf{p}_2+\textbf{p}_3-\textbf{P}_A)\\
		 &\times\Psi_{N_AL_AM_{L_A}(\textbf{p}_1\textbf{p}_2\textbf{p}_3)}\chi^{123}_{S_AM_{S_A}}|q_1(\textbf{p}_1)q_2(\textbf{p}_2)q_3(\textbf{p}_3)\rangle,
	\end{aligned}
\end{equation}
 and the wave function of the final meson state $|C\rangle$ is
\begin{equation}
		\begin{aligned}
	&|C(N^{2S_C+1}_CL_CJ_CM_{J_C})(\textbf{P}_C)\rangle\\
	=&\sqrt{2E_C}\varphi^{ab}_C\omega^{ab}_C\sum\limits_{M_{L_C}M_{S_C}}\langle L_CM_{L_C};S_CM_{S_C}|J_CM_{J_C}\rangle\\
	&\times\int d^3\textbf{p}_ad^3\textbf{p}_b\delta^3(\textbf{p}_a+\textbf{p}_b-\textbf{P}_C)\\
	&\times\Psi_{N_CL_CM_{L_C}(\textbf{p}_a\textbf{p}_b)}\chi^{ab}_{S_CM_{S_C}}|q_a(\textbf{p}_a)q_b(\textbf{p}_b)\rangle.	
		\end{aligned}
\end{equation}
The $\textbf{p}_i$ ($i = 1, 2, 3$ and $a, b$) stands for the momentum of quarks
in hadron $|A\rangle$ and $|C\rangle$. $\textbf{P}_A$ and $\textbf{P}_C$ denotes the momentum of initial baryon $|A\rangle$ and final meson $|C\rangle$, respectively.
The spatial wave functions of baryons and mesons are described with simple harmonic oscillator wave functions. Thus, the spatial wave function of  baryon $|A\rangle$ or $|B\rangle$ without the radial excitation reads
\begin{equation}
\begin{split}
\psi(l_{\rho},m_{\rho},l_{\lambda},m_{\lambda})~~~~~~~~~~~~~~~~~~~~~~~~~~~~~~~~~~~~~~~~~~~~~~~~~~~~~~~~~~~~~~~~~~~~~~~~~\\
=3^{\frac{3}{4}}(-i)^{l_{\rho}}\left[\dfrac{2^{l_{\rho}+2}}{\sqrt{\pi}(2l_{\rho}+1)!!}\right]^\frac{1}{2}
\left(\dfrac{1}{\alpha}\right)^{l_{\rho}+\frac{3}{2}}{\rm {exp}}(-\frac{\textbf{p}_{\rho}^2}{2\alpha_{\rho}^2})\mathcal{Y}^{m_{\rho}}_{l_{\rho}}(\textbf{p}_{\rho})\\
\times(-i)^{l_{\lambda}}\left[\dfrac{2^{l_{\lambda}+2}}{\sqrt{\pi}(2l_{\lambda}+1)!!}\right]^\frac{1}{2}
\left(\dfrac{1}{\alpha}\right)^{l_{\lambda}+\frac{3}{2}}{\rm {exp}}(-\frac{\textbf{p}_{\lambda}^2}{2\alpha_{\lambda}^2})\mathcal{Y}^{m_{\lambda}}_{l_{\lambda}}(\textbf{p}_{\lambda}),
\end{split}
\end{equation}
where $\textbf{p}_{\rho}$ denotes the relative momentum within the light diquark, and $\textbf{p}_{\lambda}$ is the relative momentum between the light diquark and the heavy quark in the baryon.
The ground state spatial wave function of  meson $|C\rangle$ is
\begin{equation}
	\psi_{0,0}=\left(\dfrac{R^2}{\pi}\right)^\frac{3}{4}{\rm {exp}}\left(-\dfrac{R^2\textbf{p}^2_{ab}}{2}\right).
\end{equation}
$\textbf{p}_{ab}$ stands for the relative momentum between the
quark and antiquark in the meson. Then, the partial decay amplitude in the center of mass frame can be obtained by
\begin{equation}
	\begin{aligned}
		&\mathcal{M}^{M_{J_A}M_{J_B}M_{J_C}}(A\rightarrow B+C)\\
		=&\gamma\sqrt{8E_AE_BE_C}\prod_{A,B,C}\langle\chi^{124}_{S_BM_{S_B}}\chi^{35}_{S_CM_{S_C}}|\chi^{123}_{S_AM_{S_A}}\chi^{45}_{1-m}\rangle\\
		&\langle\varphi^{124}_B\varphi^{35}_C|\varphi^{123}_A\varphi^{45}_0\rangle I^{M_{L_A},m}_{M_{L_B},M_{L_C}}(\textbf{p}).
	\end{aligned}
\end{equation}
Here, $I^{M_{L_A},m}_{M_{L_B},M_{L_C}}(\textbf{p})$ denotes the spatial integration, and $\prod_{A,B,C}$ is the Clebsch-Gorden coefficients, which  can be expanded as
\begin{equation}
	\begin{aligned}
			\sum&\langle L_BM_{L_B};S_BM_{S_B}|J_B,M_{J_B}\rangle \langle L_CM_{L_C};S_CM_{S_C}|J_C,M_{J_C}\rangle\\
			 &\times \langle L_AM_{L_A};S_AM_{S_A}|J_A,M_{J_A}\rangle \langle1m;1-m|00\rangle.\\
	\end{aligned}
\end{equation}
Finally, the decay width of initial state $|A\rangle$ decaying into final state $|BC\rangle$ can be calculated by the following formula,
\begin{equation}
\Gamma(A\rightarrow BC)=\pi^2 \dfrac{|\textbf{p}|}{M^2_A}\dfrac{1}{2J_A+1}\sum_{M_{J_A},M_{J_B},M_{J_C}}|\mathcal{M}^{M_{J_A},M_{J_B},M_{J_C}}|^2.
\end{equation}
Here, $\textbf{p}$ is the momentum of the daughter baryon $|B\rangle$ in the center of mass frame of baryon $|A\rangle$, which can be obtained by
\begin{equation}
	|\textbf{p}|=\dfrac{\sqrt{[M_A^2-(M_B-M_C)^2][M_A^2-(M_B+M_C)^2]}}{2M_A}.
\end{equation}

In our calculation, we adopt $m_u$ = $m_d$ = 330 MeV , $m_s$ = 450 MeV , and $m_c$ = 1700 MeV for the constituent quark mass. The harmonic oscillator strength $R$ for light flavor mesons is taken to be $R=2.5$ $\rm{GeV^{-1}}$, and that for $D$ meson is $R=1.67$ $\rm{GeV^{-1}}$~\cite{Godfrey:2015dva}. The parameter of the $\rho$-mode excitation between the two light quarks is taken as $\alpha_{\rho}$=0.4 GeV. The other parameter $\alpha_\lambda$ is obtained by the relation~\cite{Zhong:2007gp}
\begin{equation}
	\alpha_\lambda=\left(\dfrac{3m_Q}{2m_q+m_Q}\right)^\frac{1}{4}\alpha_\rho.
\end{equation}
The value of vacuum pair-production strength $\gamma$ is fixed as $\gamma=$6.95, which is the same as that in our previous work~\cite{Xiao:2017dly}. Additionally, the masses of final baryons and mesons are derived from Particle Data Group~\cite{ParticleDataGroup:2020ssz}.

\subsection{Chiral quark model}

Within the chiral quark model (ChQM), the low energy effective quark-pseudoscalar-meson coupling at tree level in the SU(3) flavor basis reads
\cite{Manohar:1983md}
\begin{eqnarray}\label{STcoup}
H_m=\sum_j
\frac{1}{f_m}\bar{\psi}_j\gamma^{j}_{\mu}\gamma^{j}_{5}\psi_j\vec{\tau}\cdot\partial^{\mu}\vec{\phi}_m.
\end{eqnarray}
Here, $f_m$ denotes the pseudoscalar meson decay constant and $\psi_j$ represents the $j$-th quark field in a baryon.
$\phi_m$ is the
pseudoscalar meson octet, which is
 \begin{eqnarray}
\phi_m=
\begin{pmatrix}
\frac{1}{\sqrt{2}}\pi^0+\frac{1}{\sqrt{6}}\eta & \pi^+ & K^+ \cr
\pi^- & -\frac{1}{\sqrt{2}}\pi^0+\frac{1}{\sqrt{6}}\eta & K^0 \cr
K^- & \bar{K}^0 & -\sqrt{\frac{2}{3}\eta}
\end{pmatrix}.
\end{eqnarray}
In this work, because of the harmonic oscillator spatial wave function of baryons being nonrelativistic form, the coupling is adopt the nonrelativistic form as well and is written as \cite{Li:1994cy,Li:1997gd,Zhao:2002id}
\begin{eqnarray}\label{non-relativistic-expansST}
H^{nr}_{m}=\sum_j\Big\{\frac{\omega_m}{E_f+M_f}\vsig_j\cdot\textbf{P}_f+ \frac{\omega_m}{E_i+M_i}\vsig_j \cdot\textbf{P}_i\\  \nonumber
-\vsig_j \cdot \textbf{q} +\frac{\omega_m}{2\mu_q}\vsig_j\cdot\textbf{p}'_j\Big\}I_j \phi_m.
\end{eqnarray}
In this equation, $\textbf{q}$ and $\omega_m$ denote the three-vector momentum and energy
of the meson, respectively; The $\vsig_j$ represents the Pauli spin vector; $M_{i(f)}$, $E_{i(f)}$, and $\textbf{P}_{i(f)}$ stand for the mass, energy and three-vector momentum of the initial (final) baryon; $\mu_q$ is
the reduced mass of the $j$-th quark in the initial and final
baryons. $\textbf{p}'_j=\textbf{p}_j-(m_j/M) \textbf{P}_{c.m.}$ stands for the
internal momentum of the $j$-th quark in the baryon rest frame. $I_j$ is the isospin operator associated with the pseudoscalar meson and $\varphi_m=e^{-i\textbf{q}\cdot\textbf{r}_j}$ is the plane wave for the emitting
light pseudoscalar meson from a decay process.

Hence, the partial decay amplitudes $M_{J_{iz},J_{fz}}$ of a light pseudoscalar meson emission in a baryon's strong decays can be calculated. Then, the strong decay width can be worked out by
\begin{equation}
\Gamma=\left(\frac{\delta}{f_m}\right)^2\frac{(E_f +M_f)|q|}{4\pi
	M_i}\frac{1}{2J_i+1}\sum_{J_{iz}J_{fz}}^{}|M_{J_{iz},J_{fz}}|^2,
\end{equation}
where $J_{iz}$ and $J_{fz}$ denote the third components of the total angular momenta of the initial and final baryons, respectively. $\delta$ is a global parameter accounting for the strength of the quark-meson couplings. Here, we fix its value the
same as that in Refs.\cite{Zhong:2007gp,Zhong:2008kd}, i.e. $\delta=0.557$ MeV, which is determined by experimental data. The decay constants $f_{m}$ for $\pi$ and $K$ mesons
are taken as $f_{\pi}=132$ MeV and $f_{K}=160$ MeV, respectively. In addition, the constituent quark masses are the same as in that of the quark pair creation model.

\subsection{Coupling scheme}

It should be pointed out that due to the heavy quark symmetry, the physical states may be closer to the $j$-$j$ coupling scheme.
In the heavy quark symmetry limit, the states within the $j$-$j$ coupling scheme are constructed by
\begin{equation}
|J^P,j\rangle=	\left|\left\{[(l_{\rho}l_{\lambda})_L s_{\rho}]_j s_Q\right\}_{J^P}
	\right \rangle.
\end{equation}
In the expression, $l_{\rho}$ and $l_{\lambda}$ are the quantum numbers of the orbital angular momentum for
$\rho$- and $\lambda$-mode excitations, respectively. The the quantum number of total orbital angular momentum
$L = |l_{\rho}-l_{\lambda}|,\cdots , l_{\rho}+l_{\lambda} $. $s_{\rho}$ is the spin quantum number of
the light quark pair, and $s_Q$ is the spin quantum number of the
heavy quark.

The states within the the
$j$-$j$ coupling scheme can be expressed as linear combinations
of the states within the $L$-$S$ coupling via~\cite{Roberts:2007ni}
\begin{equation}
\begin{aligned}
	\left|\left\{[(l_{\rho}l_{\lambda})_L s_{\rho}]_j s_Q\right\}_{J^P}
	\right \rangle&=(-1)^{L+s_{\rho}+\frac{1}{2}+J}\sqrt{2J+1}\sum_{S}\sqrt{2S+1}\\
	&\begin{pmatrix}
		L&s_{\rho}&j\\s_Q&J&S
	\end{pmatrix}	\left|\left\{[(l_{\rho}l_{\lambda})_L (s_{\rho}s_Q)_S]_J \right\}
\right\rangle.
\end{aligned}
\end{equation}
The quantum number of total spin angular momentum $S =|s_{\rho}-s_Q|,\cdots , s_{\rho}+s_Q $.
$J$ is the quantum number of total angular momentum.

\section{Calculations and Results }

We calculate the strong decays of
 $1P$-wave and $1D$-wave excited $\Sigma_c$ baryons within the $j$-$j$ coupling scheme
in the framework of the QPC model. Both $\lambda$-mode and $\rho$-mode excitations are considered in the present work.
In addition, for a comparison we also give the decay properties of those $\Sigma_c$ excitations within the ChQM.
We hope our calculations being helpful for searching those missing $\Sigma_c$ states in forthcoming experiments.

\subsection{$1P$-wave $\lambda$-mode excitations}

For the $1P$-wave $\lambda$-mode $\Sigma_c$ baryons, there are five states according to the quark model classification, which are $\Sigma_c|J^P=1/2^-,0\rangle_{\lambda}$, $\Sigma_c|J^P=1/2^-,1\rangle_{\lambda}$, $\Sigma_c|J^P=3/2^-,1\rangle_{\lambda}$, $\Sigma_c|J^P=3/2^-,2\rangle_{\lambda}$ and $\Sigma_c|J^P=5/2^-,2\rangle_{\lambda}$. They have not been established in experiments. Their theoretical masses and possible two-body decay channels are listed in Table~\ref{table1}.

\begin{figure}[]
	\centering \epsfxsize=8 cm \epsfbox{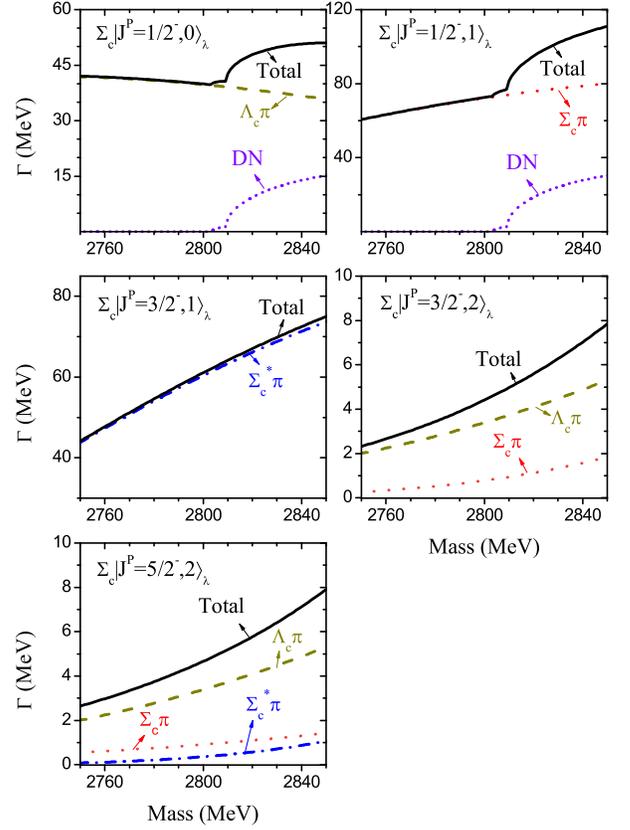}
	\caption{ Partial and total strong decay widths of the $1P$-wave $\lambda$-mode $\Sigma_c$ excitations as functions of their masses. Some decay channels are too small to show in figure.}
    \label{FIG2}
\end{figure}

From the table, it is known that the predicted masses of the $1P$-wave $\lambda$-mode $\Sigma_c$ baryons are about $M\simeq(2750-2850)$ MeV, and the possible two-body decay channels are $\Sigma_c^{(*)}\pi$, $\Lambda_c\pi$ and $DN$. Adopting the predicted masses from Ref.~\cite{Yu:2022ymb}, we calculate their decay properties within the QPC model, the results are collected in Table~\ref{Plambda}. In addition, considering the uncertainty of the mass predictions of the $\lambda$-mode $1P$-wave $\Sigma_c$ states, we plot the strong decay properties as a function of the mass in the region of $M=(2750-2850)$ MeV in Fig.~\ref{FIG2} with the QPC model. From the figure, it is found that the decay properties of the five $1P$-wave $\lambda$-mode $\Sigma_c$ excitations are sensitive to the masses, especially the two $J^P=1/2^-$ states $\Sigma_c|J^P=1/2^-,0\rangle_{\lambda}$ and $\Sigma_c|J^P=1/2^-,1\rangle_{\lambda}$. When the masses of $\Sigma_c|J^P=1/2^-,0\rangle_{\lambda}$ and $\Sigma_c|J^P=1/2^-,1\rangle_{\lambda}$ are above the threshold of $DN$, their decay rates into $DN$ will be significant. Hence, the two $J^P=1/2^-$ states may be explored in the $DN$ channel in experiments. In Ref.~\cite{Wang:2021bmz}, the decay
properties were also studied within the ChQM, for a comparison, the results are listed in Table~\ref{Plambda}.

\begin{table*}[]
	\caption{\label{Plambda}The strong decay properties of the $\lambda$-mode $1P$-wave $\Sigma_c$ states, which masses are taken from the predictions in Ref.~\cite{Yu:2022ymb}. $\Gamma_{\text{Total}}$ stands for the total decay width. The unit of the width and mass is MeV.}
	\centering
	\begin{tabular}{ccccccccccccccccc}	
		\hline\hline
		&\multicolumn{2}{c}{$\underline{~~\Sigma_{c}\ket{J^{P}=\frac{1}{2}^{-},0}_{\lambda}~~}$} &\multicolumn{2}{c}{$\underline{~~\Sigma_{c}\ket{J^{P}=\frac{1}{2}^{-},1}_{\lambda}~~}$}
&\multicolumn{2}{c}{$\underline{~~\Sigma_{c}\ket{J^{P}=\frac{3}{2}^{-},1}_{\lambda}~~}$}&\multicolumn{2}{c}{$\underline{~~\Sigma_{c}\ket{J^{P}=\frac{3}{2}^{-},2}_{\lambda}~~}$}
&\multicolumn{2}{c}{$\underline{~~\Sigma_{c}\ket{J^{P}=\frac{5}{2}^{-},2}_{\lambda}~~}$}\\
&\multicolumn{2}{c}{$M$=2823}&\multicolumn{2}{c}{$M$=2809}&\multicolumn{2}{c}{$M$=2829}&\multicolumn{2}{c}{$M$=2802}&\multicolumn{2}{c}{$M$=2835}\\
				\cline{2-11}
Decay width&QPC&ChQM~\cite{Wang:2021bmz}&QPC&ChQM~\cite{Wang:2021bmz}&QPC&ChQM~\cite{Wang:2021bmz}&QPC&ChQM~\cite{Wang:2021bmz}&QPC&ChQM~\cite{Wang:2021bmz}\\
		\hline				
		$\Gamma[DN]$&10.3&-&2.7&-&...&-&...&-&...&-\\
		$\Gamma[\Sigma_{c}\pi]$&0.1&...&73.9&30.2&1.0&5.5&0.8&6.6&1.3&4.8\\
		$\Gamma[\Lambda_c\pi]$&38.1&5.7&...&...&...&...&3.5&29.8&4.7&38.6\\
		$\Gamma[\Sigma^{*}_c\pi]$&...&...&0.3&0.5&68.5&31.2&0.2&2.0&0.8&2.1\\
		$\Gamma_{\text{Total}}$&48.5&5.7&76.9&30.7&69.5&36.7&4.5&38.4&6.8&45.5\\
		\hline \hline
\end{tabular}	
\end{table*}


The $\Sigma_c|J^P=1/2^-,0\rangle_{\lambda}$ state mainly decays into $\Lambda_c\pi$. Within the QPC model,
this state is a moderate width state with a width of $\Gamma\simeq 49$ MeV. If the mass is taken as the recent quark model prediction $M=2823$ MeV~\cite{Yu:2022ymb},
the $\Sigma_c|J^P=1/2^-,0\rangle_{\lambda}$ state may has a significant decay rate into $DN$, the branching fraction is predicted to be
\begin{eqnarray}
\frac{\Gamma[\Sigma_{c}\ket{J^{P}=\frac{1}{2}^{-},0}_{\lambda}\rightarrow DN]}{\Gamma_{\text{Total}}}\sim21\%.
\end{eqnarray}
However, within the ChQM, the $\Sigma_c|J^P=1/2^-,0\rangle_{\lambda}$ state might be a narrow state,
the predicted partial width of the $\Lambda_c\pi$ channel, $\Gamma[\Sigma_{c}\ket{J^{P}=\frac{1}{2}^{-},0}_{\lambda}\to \Lambda_c\pi]\simeq 5.7$ MeV,
is about one order of magnitude smaller than that predicted with the QPC model.
The decay properties of $\Sigma_c|J^P=1/2^-,0\rangle_{\lambda}$ predicted within the QPC model are in good agreement with the nature
of $\Sigma_c(2800)$ observed in the $\Lambda_c\pi$ channel. It should be mentioned that according to the equal spacing rule, the mass of
$\Sigma_c|J^P=1/2^-,0\rangle_{\lambda}$ is predicted to be about $2755$ MeV~\cite{Wang:2021bmz},
which is below the $DN$ mass threshold. If the observed state $\Sigma_c(2800)$ corresponds to $\Sigma_c|J^P=1/2^-,0\rangle_{\lambda}$ indeed and its mass is above the threshold of $DN$, besides the $\Lambda_c\pi$ channel, the $DN$ may be the other interesting channel for the observation of $\Sigma_c(2800)$ in future experiments.

The $\Sigma_c|J^P=1/2^-,1\rangle_{\lambda}$ and $\Sigma_c|J^P=3/2^-,1\rangle_{\lambda}$ states,
dominantly decay into $\Sigma_c\pi$ and $\Sigma_c^*\pi$ channels, respectively. In the QPC model,
they are predicted to be moderate width states with a comparable width of $\Gamma\sim 70$ MeV, which is a factor 2 larger
than that of the ChQM. The branching fractions of their dominant decay channel can reach up to
\begin{eqnarray}
\frac{\Gamma[\Sigma_{c}\ket{J^{P}=\frac{1}{2}^{-},1}_{\lambda}\rightarrow \Sigma_c\pi]}{\Gamma_{\text{Total}}}\sim96\%,
\end{eqnarray}
\begin{eqnarray}
\frac{\Gamma[\Sigma_{c}\ket{J^{P}=\frac{3}{2}^{-},1}_{\lambda}\rightarrow \Sigma_c^*\pi]}{\Gamma_{\text{Total}}}\sim85-99\%.
\end{eqnarray}
The $\Sigma_c|J^P=1/2^-,1\rangle_{\lambda}$ and $\Sigma_c|J^P=3/2^-,1\rangle_{\lambda}$ states are likely to be observed in the $\Lambda_c\pi\pi$ final state via the decay chains $\Sigma_c|J^P=1/2^-(3/2^-),1\rangle_{\lambda}\rightarrow \Sigma_c\pi(\Sigma_c^*\pi)\rightarrow \Lambda_c\pi\pi$.

Furthermore, $\Sigma_c|J^P=1/2^-,1\rangle_{\lambda}$ may have a sizeable decay rate into $DN$, the predicted branching fraction is
\begin{eqnarray}
\frac{\Gamma[\Sigma_{c}\ket{J^{P}=\frac{1}{2}^{-},1}_{\lambda}\rightarrow DN]}{\Gamma_{\text{Total}}}\sim4\%.
\end{eqnarray}
Hence, $DN$ may be a good channel for looking for the missing state $\Sigma_c|J^P=1/2^-,1\rangle_{\lambda}$.

The other two $1P$-wave $\lambda$-mode states $\Sigma_c|J^P=3/2^-,2\rangle_{\lambda}$ and $\Sigma_c|J^P=5/2^-,2\rangle_{\lambda}$
mainly decay into the $\Lambda_c\pi$ channel. The predicted decay widths are very different between the QPC model
and the ChQM. Within the QPC model, both $\Sigma_c|J^P=3/2^-,2\rangle_{\lambda}$ and
$\Sigma_c|J^P=5/2^-,2\rangle_{\lambda}$ are most likely to be narrow states with a width of about several MeV,
while according to the ChQM predictions, they may have a moderate width of $\Gamma\sim40$ MeV.
The predicted decay properties within the ChQM are consistent with the observations of
$\Sigma_c(2800)$. Thus, in Ref.~\cite{Wang:2021bmz}, $\Sigma_c(2800)$
is suggested to be a structure potentially arising from two overlapping $1P$-wave $\Sigma_c$ states with $J^P=3/2^-$
and $J^P=5/2^-$. Due to the extremely similar decay properties of $\Sigma_c|J^P=3/2^-,2\rangle_{\lambda}$ and $\Sigma_c|J^P=5/2^-,2\rangle_{\lambda}$, it will be a big challenge to distinguish them from each other in experiments.

\subsection{$1P$-wave $\rho$-mode excitations}

For the $1P$-wave $\rho$-mode $\Sigma_c$ baryons, there are two states $\Sigma_c|J^P=1/2^-,1\rangle_{\rho}$ and $\Sigma_c|J^P=3/2^-,1\rangle_{\rho}$. For their masses, there are a few discussions in theoretical references and we have collected in Table~\ref{table1} as well. From the table, the masses of the $1P$-wave $\rho$-mode $\Sigma_c$ excitations are about $M\sim$2.9 GeV. Meanwhile, we notice that the masses of the $1P$-wave $\rho$-mode $\Sigma_c$ baryons are above the threshold of $\Lambda_c\pi$ and $DN$, while their strong decays are forbidden due to the orthogonality of spatial wave functions when we adopt the simple harmonic oscillator wave functions in this work. Hence, we mainly focus on their strong decays into $\Sigma_c\pi$ and $\Sigma^*_c\pi$.


\begin{table}[]
	\caption{\label{1Prho}Partial decay widths(MeV) and branching fractions for the $1P$-wave $\rho$-mode $\Sigma_c$ states. The numbers in parentheses stand for the corresponding masses(MeV).}
	\centering
	\begin{tabular}{l|cc|cc|cc|cc}
		\hline \hline
		\multirow{3}{*}{ }&\multicolumn{4}{c}{$\Sigma_{c}\ket{J^{P}=\frac{1}{2}^{-},1}_{\rho}$(2909)}
		~~~& \multicolumn{4}{c}{$\Sigma_{c}\ket{J^{P}=\frac{3}{2}^{-},1}_{\rho}$(2910)}\\
		\cline{2-9}
		&\multicolumn{2}{c}{QPC}&\multicolumn{2}{c}{ChQM}&\multicolumn{2}{c}{QPC}&\multicolumn{2}{c}{ChQM}\\
		\cline{2-9}
		&$\Gamma_i$ & $B_i(\%)$&$\Gamma_i$ & $B_i(\%)$&$\Gamma_i$ & $B_i(\%)$&$\Gamma_i$ & $B_i(\%)$\\
		\hline
		$\Sigma_{c}$$\pi$& 174.3&92&50&66&18.8&12&27.5&30\\
		$\Sigma^{*}_{c}$$\pi$&16.0&8 &25.5&34&136.0&88&63.5&70\\
		\hline
		Total&190.3&&75.5& &154.8&&91.0&\\
		\hline\hline
\end{tabular}
\end{table}

Fixing the masses of the $1P$-wave $\rho$-mode $\Sigma_c$ excitations on the predicted masses from Ref.~\cite{Yoshida:2015tia}, we discuss their decay properties within the QPC model. The results are listed in Table~\ref{1Prho}. It is found that the two $1P$-wave $\rho$-mode $\Sigma_c$ excitations may be broad states with a total decay width of about $\Gamma\sim(150-200)$ MeV. The $\Sigma_c|J^P=1/2^-,1\rangle_{\rho}$ is mostly saturated by the decay channel $\Sigma_c\pi$ and the branching fraction for the $\Sigma_c\pi$ channel is
\begin{eqnarray}
\frac{\Gamma[\Sigma_{c}\ket{J^{P}=\frac{1}{2}^{-},1}_{\rho}\rightarrow \Sigma_c\pi]}{\Gamma_{\text{Total}}}\sim92\%.
\end{eqnarray}
While, the decay of $\Sigma_{c}\ket{J^{P}=\frac{3}{2}^{-},1}_{\rho}$ is governed by $\Sigma^*_c\pi$ with branching fraction
\begin{eqnarray}
\frac{\Gamma[\Sigma_{c}\ket{J^{P}=\frac{3}{2}^{-},1}_{\rho}\rightarrow \Sigma^*_c\pi]}{\Gamma_{\text{Total}}}\sim88\%.
\end{eqnarray}

\begin{figure}[h]
	\centering \epsfxsize=8.5 cm \epsfbox{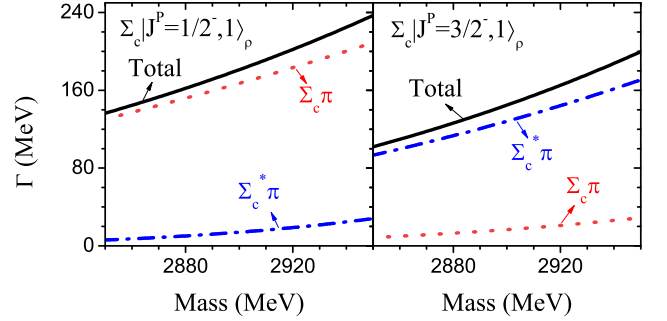}
	\caption{ Partial and total strong decay widths of the
		$1P$-wave $\rho$-mode $\Sigma_c$ states. }
\label{FIG3}
\end{figure}

For a comparison, the predicted results with the ChQM are collected in Table~\ref{1Prho} as well. It is found that the
predicted main decay channels with the two models are agreement to each other, however, the predicted total decay widths
with the QPC model are about a factor of 2 larger than that with the ChQM.

Similarly, the predicted masses of the $1P$-wave $\rho$-mode $\Sigma_c$ excitations certainly have a large uncertainty, which may bring uncertainties to the theoretical results. To investigate this effect, we analyze the decay properties of the $1P$-wave $\rho$-mode $\Sigma_c$ excitations as a function of the mass in Fig.~\ref{FIG3} with the QPC model. Within the mass varying in the region of $M=(2850-2950)$ MeV, the total widths of $\Sigma_c|J^P=1/2^-,1\rangle_{\rho}$ and $\Sigma_c|J^P=3/2^-,1\rangle_{\rho}$ are $\Gamma>100$ MeV, which shows some
sensitivities to the mass changing. The $\rho$-mode states $\Sigma_c|J^P=1/2^-,1\rangle_{\rho}$ and $\Sigma_c|J^P=3/2^-,1\rangle_{\rho}$
are much broader than the $1P$-wave $\lambda$-mode states, thus, they may be difficult to be observed in experiments.

\subsection{$1D$-wave $\lambda$-mode excitations}

For the $1D$-wave $\lambda$-mode $\Sigma_c$ baryons, there are six states $\Sigma_c|J^P=1/2^+,1\rangle_{\lambda\lambda}$, $\Sigma_c|J^P=3/2^+,1\rangle_{\lambda\lambda}$, $\Sigma_c|J^P=3/2^+,2\rangle_{\lambda\lambda}$, $\Sigma_c|J^P=5/2^+,2\rangle_{\lambda\lambda}$, $\Sigma_c|J^P=5/2^+,3\rangle_{\lambda\lambda}$ and $\Sigma_c|J^P=7/2^+,3\rangle_{\lambda\lambda}$. According to the theoretical predictions by various methods, the mass of the $1D$-wave $\lambda$-mode $\Sigma_c$ baryons is about $M\sim3.0$ GeV. Fixing their masses at the predictions in Ref.~\cite{Ebert:2011kk}, we calculate their strong decay properties with both the QPC model and ChQM, the results
are listed in Table~\ref{Dlambda}.

\begin{table*}[]
	\caption{\label{Dlambda}The strong decay properties of the $\lambda$-mode $1D$-wave $\Sigma_c$ states, which masses are taken from the predictions in Ref.~\cite{Ebert:2011kk}. $\Gamma_{\text{Total}}$ stands for the total decay width. The unit of the width and mass is MeV.}
	\centering
	\begin{tabular}{ccccccccccccccccc}	
		\hline\hline
		&\multicolumn{2}{c}{$\underline{~\Sigma_{c}\ket{J^{P}=\frac{1}{2}^{+},1}_{\lambda\lambda}~}$} &\multicolumn{2}{c}{$\underline{~\Sigma_{c}\ket{J^{P}=\frac{3}{2}^{+},1}_{\lambda\lambda}~}$}
&\multicolumn{2}{c}{$\underline{~\Sigma_{c}\ket{J^{P}=\frac{3}{2}^{+},2}_{\lambda\lambda}~}$}&\multicolumn{2}{c}{$\underline{~\Sigma_{c}\ket{J^{P}=\frac{5}{2}^{+},2}_{\lambda\lambda}~}$}
&\multicolumn{2}{c}{$\underline{~\Sigma_{c}\ket{J^{P}=\frac{5}{2}^{+},3}_{\lambda\lambda}~}$}&\multicolumn{2}{c}{$\underline{~\Sigma_{c}\ket{J^{P}=\frac{7}{2}^{+},3}_{\lambda\lambda}~}$}\\		 &\multicolumn{2}{c}{$M$=3041}&\multicolumn{2}{c}{$M$=3043}&\multicolumn{2}{c}{$M$=3040}&\multicolumn{2}{c}{$M$=3038}&\multicolumn{2}{c}{$M$=3023}&\multicolumn{2}{c}{$M$=3013}\\
		\cline{2-13}
Decay width		&QPC&ChQM&QPC&ChQM&QPC&ChQM&QPC&ChQM&QPC&ChQM&QPC&ChQM\\
		\hline
		$\Gamma[DN]$&17.1&-&1.1&-&9.6&-&0.3&-&...&-&0.6&-\\
        $\Gamma[D^*N]$&3.9&-&7.5&-&4.6&-&6.8&-&0.1&-&...&-\\
		$\Gamma[\Sigma_{c}\pi]$&2.5&3.5&0.7&0.9&5.3&7.9&0.3&4.8&0.3&4.7&0.1&2.4\\
		$\Gamma[\Lambda_c\pi]$&2.8&2.3&2.4&2.2&0.1&...&...&...&0.7&15.5&0.6&14.5\\
		$\Gamma[\Sigma^{*}_c\pi]$&1.0&1.7&2.8&4.3&3.5&6.0&5.5&10.8&0.2&5.2&0.2&1.5\\
        $\Gamma[\Xi_cK]$&0.9&3.4&0.8&3.6&...&...&...&...&...&...&...&...\\
        $\Gamma[\Lambda_c\ket{J^P=\frac{1}{2}^{-},1}_{\lambda}\pi]$&37.2&13.8&...&8.1&0.4&0.2&0.2&...&1.7&2.2&...&...\\
        $\Gamma[\Lambda_c\ket{J^P=\frac{3}{2}^{-},1}_{\lambda}\pi]$&0.1&1.4&37.3&10.8&0.3&0.2&0.5&1.9&0.3&1.2&1.0&8.0\\
        $\Gamma[\Sigma_c\ket{J^P=\frac{1}{2}^{-},0}_{\lambda}\pi]$&4.3&0.3&...&0.4&0.2&1.3&...&0.5&...&0.2&1.0&...\\
        $\Gamma[\Sigma_c\ket{J^P=\frac{1}{2}^{-},1}_{\lambda}\pi]$&27.9&24.9&0.3&0.5&0.5&4.3&0.2&0.1&...&3.9&0.8&...\\
        $\Gamma[\Sigma_c\ket{J^P=\frac{3}{2}^{-},1}_{\lambda}\pi]$&...&3.4&6.7&0.6&4.0&2.9&2.0&0.1&2.3&...&2.7&0.1\\
        $\Gamma[\Sigma_c\ket{J^P=\frac{3}{2}^{-},2}_{\lambda}\pi]$&...&1.5&4.6&0.3&55.6&22.9&0.5&...&1.0&0.3&0.5&0.2\\
        $\Gamma[\Sigma_c\ket{J^P=\frac{5}{2}^{-},2}_{\lambda}\pi]$&14.7&0.9&3.3&1.3&5.0&0.9&8.6&7.6&6.9&0.8&1.4&0.4\\
        $\Gamma_{\text{Total}}$&112.4&57.0&67.5&33.0&89.1&46.6&24.9&25.8&13.5&34.0&8.9&27.1\\
		\hline \hline
&\multicolumn{2}{c}{$\underline{~\Sigma_{c}\ket{J^{P}=\frac{1}{2}^{+},1}_{\rho\rho}~}$} &\multicolumn{2}{c}{$\underline{~\Sigma_{c}\ket{J^{P}=\frac{3}{2}^{+},1}_{\rho\rho}~}$}
&\multicolumn{2}{c}{$\underline{~\Sigma_{c}\ket{J^{P}=\frac{3}{2}^{+},2}_{\rho\rho}~}$}&\multicolumn{2}{c}{$\underline{~\Sigma_{c}\ket{J^{P}=\frac{5}{2}^{+},2}_{\rho\rho}~}$}
&\multicolumn{2}{c}{$\underline{~\Sigma_{c}\ket{J^{P}=\frac{5}{2}^{+},3}_{\rho\rho}~}$}&\multicolumn{2}{c}{$\underline{~\Sigma_{c}\ket{J^{P}=\frac{7}{2}^{+},3}_{\rho\rho}~}$}\\
		 &\multicolumn{2}{c}{$M$=3141}&\multicolumn{2}{c}{$M$=3143}&\multicolumn{2}{c}{$M$=3140}&\multicolumn{2}{c}{$M$=3138}&\multicolumn{2}{c}{$M$=3123}&\multicolumn{2}{c}{$M$=3113}\\
		\cline{2-13}
Decay width		&QPC&ChQM&QPC&ChQM&QPC&ChQM&QPC&ChQM&QPC&ChQM&QPC&ChQM\\
		\hline
		$\Gamma[\Sigma_{c}\pi]$&69.0&14.1&29.0&3.5&126.4&31.7&11.1&10.2&11.1&10.4&5.7&5.4\\
		$\Gamma[\Lambda_c\pi]$&136.8&14.3&137.8&14.2&...&...&...&...&21.7&24.3&20.3&23.1\\
		$\Gamma[\Sigma^{*}_c\pi]$&24.8&6.9&62.6&17.3&33.3&17.3&136.3&39.4&7.4&18.8&8.9&9.3\\
        $\Gamma[\Xi_cK]$&23.1&19.6&37.5&19.9&...&...&...&...&1.0&0.6&0.8&0.5\\
        $\Gamma[\Xi^{\prime}_cK]$&1.9&2.1&0.6&0.5&3.8&4.6&...&...&...&...&...&...\\
        $\Gamma[\Lambda_c\ket{J^P=\frac{1}{2}^{-},1}_{\lambda}\pi]$&9.2&1.1&1.2&1.5&...&...&...&...&1.7&0.6&...&0.5\\
        $\Gamma[\Lambda_c\ket{J^P=\frac{3}{2}^{-},1}_{\lambda}\pi]$&1.8&1.9&8.8&2.4&...&...&...&...&0.4&0.8&1.8&0.6\\

        $\Gamma[\Sigma_c\ket{J^P=\frac{1}{2}^{-},0}_{\lambda}\pi]$&...&0.2&...&0.1&...&0.1&0.1&0.1&...&...&...&...\\
        $\Gamma[\Sigma_c\ket{J^P=\frac{1}{2}^{-},1}_{\lambda}\pi]$&7.0&0.1&0.2&0.1&1.6&0.2&0.7&...&1.3&0.1&...&...\\         $\Gamma[\Sigma_c\ket{J^P=\frac{3}{2}^{-},1}_{\lambda}\pi]$&1.3&0.2&2.1&...&1.3&...&0.2&0.1&0.1&...&0.2&...\\
        $\Gamma[\Sigma_c\ket{J^P=\frac{3}{2}^{-},2}_{\lambda}\pi]$&...&0.4&0.1&0.5&6.5&0.1&0.1&...&0.2&...&0.1&...\\
        $\Gamma[\Sigma_c\ket{J^P=\frac{5}{2}^{-},2}_{\lambda}\pi]$&1.9&0.1&0.8&...&1.1&0.1&5.5&0.5&0.1&...&0.3&...\\
        $\Gamma_{\text{Total}}$&276.7&61.0&280.7&60.0&174.0&54.1&153.9&50.3&44.9&55.6&38.1&39.4\\
		\hline \hline
\end{tabular}	
\end{table*}

For the $\Sigma_c|J^P=1/2^+,1\rangle_{\lambda\lambda}$, similar decay properties are predicted within the two models.
The $\Sigma_c|J^P=1/2^+,1\rangle_{\lambda\lambda}$ may have a width of about 10s-100 MeV.
The main decay channels of $\Sigma_c|J^P=1/2^+,1\rangle_{\lambda\lambda}$ are $\Lambda_c|J^P=1/2^-,1\rangle_{\lambda}\pi$ and $\Sigma_c|J^P=1/2^-,1\rangle_{\lambda}\pi$.
With the QPC model, the branching fractions are predicted to be
\begin{eqnarray}
\frac{\Gamma[\Sigma_{c}\ket{J^{P}=\frac{1}{2}^{+},1}_{\lambda\lambda}\rightarrow \Lambda_c|J^P=\frac{1}{2}^{-},1\rangle_{\lambda}\pi]}{\Gamma_{\text{Total}}}\sim33\%,
\end{eqnarray}
\begin{eqnarray}
\frac{\Gamma[\Sigma_{c}\ket{J^{P}=\frac{1}{2}^{+},1}_{\lambda\lambda}\rightarrow \Sigma_c|J^P=\frac{1}{2}^{-},1\rangle_{\lambda}\pi]}{\Gamma_{\text{Total}}}\sim25\%.
\end{eqnarray}
The large branching ratios indicate the state $\Sigma_c|J^P=1/2^+,1\rangle_{\lambda\lambda}$ is very likely be observed in the $\Sigma_c\pi\pi$ final state via the decay chains $\Sigma_c|J^P=1/2^+,1\rangle_{\lambda\lambda}\rightarrow \Lambda_c(\Sigma_c)|J^P=1/2^-,1\rangle_{\lambda}\pi\rightarrow \Sigma_c\pi\pi$.

Meanwhile, according to the QPC model prediction, the state $\Sigma_c|J^P=1/2^+,1\rangle_{\lambda\lambda}$
may have a significant decay rate into $DN$ with a branching fraction
\begin{eqnarray}
\frac{\Gamma[\Sigma_{c}\ket{J^{P}=\frac{1}{2}^{+},1}_{\lambda\lambda}\rightarrow DN]}{\Gamma_{\text{Total}}}\sim15\%.
\end{eqnarray}
Hence, the $\Sigma_c|J^P=1/2^+,1\rangle_{\lambda\lambda}$ may be observed in the $DN$ decay channel.

The $\Sigma_c|J^P=3/2^+,1\rangle_{\lambda\lambda}$ state may be a moderate state.
Within the QPC model, the total decay width is predicted to be $\Gamma\sim68$ MeV,
which is about a factor of 2 larger than that predicted within the ChQM.
The $\Sigma_c|J^P=3/2^+,1\rangle_{\lambda\lambda}$ has a large decay rate into
$\Lambda_c|J^P=3/2^-,1\rangle_{\lambda}\pi$. Within the QPC model, the branching fraction
is predicted to be
\begin{eqnarray}
\frac{\Gamma[\Sigma_{c}\ket{J^{P}=\frac{3}{2}^{+},1}_{\lambda\lambda}\rightarrow \Lambda_c|J^P=\frac{3}{2}^{-},1\rangle_{\lambda}\pi]}{\Gamma_{\text{Total}}}\sim56\%.
\end{eqnarray}
Moreover, the state $\Sigma_c|J^P=3/2^+,1\rangle_{\lambda\lambda}$ may have a sizable
decay rate into $D^*N$ with a branching fraction of
\begin{eqnarray}
\frac{\Gamma[\Sigma_{c}\ket{J^{P}=\frac{3}{2}^{+},1}_{\lambda\lambda}\rightarrow D^*N]}{\Gamma_{\text{Total}}}\sim11\%.
\end{eqnarray}
Hence, besides the decay chain $\Sigma_c|J^P=3/2^+,1\rangle_{\lambda\lambda}\rightarrow \Lambda_c|J^P=3/2^-,1\rangle_{\lambda}\pi\rightarrow\Lambda_c\pi\pi\pi$, the decay chain $\Sigma_c|J^P=3/2^+,1\rangle_{\lambda\lambda}\rightarrow D^*N\rightarrow D\pi N$ may be also
the other optional decay precess to investigate the nature of $\Sigma_c|J^P=3/2^+,1\rangle_{\lambda\lambda}$.

For the other $J^P=3/2^+$ state, $\Sigma_c|J^P=3/2^+,2\rangle_{\lambda\lambda}$, with the QPC model the width
is predicted to be $\Gamma\sim89$ MeV, which is about a factor of 2 larger than that predicted
with the ChQM. This state dominantly decays into $\Sigma_c|J^P=3/2^-,2\rangle_{\lambda}\pi$.
Its predicted fraction is
\begin{eqnarray}
\frac{\Gamma[\Sigma_{c}\ket{J^{P}=\frac{3}{2}^{+},2}_{\lambda\lambda}\rightarrow \Sigma_c|J^P=\frac{3}{2}^{-},2\rangle_{\lambda}\pi]}{\Gamma_{\text{Total}}}\sim50-60\%.
\end{eqnarray}
Thanks to the large branching fraction, the state $\Sigma_c|J^P=3/2^+,2\rangle_{\lambda\lambda}$ may be observed in the $\Lambda_c\pi\pi$ decay channel via the decay chain $\Sigma_c|J^P=3/2^+,2\rangle_{\lambda\lambda}\rightarrow \Sigma_c|J^P=\frac{3}{2}^{-},2\rangle_{\lambda}\pi\rightarrow \Lambda_c\pi\pi$.

Furthermore, the  $\Sigma_c|J^P=3/2^+,2\rangle_{\lambda\lambda}$ state has a sizeable decay rate into $DN$, with the QPC model the
branching fraction is predicted to be
\begin{eqnarray}
\frac{\Gamma[\Sigma_{c}\ket{J^{P}=\frac{3}{2}^{+},2}_{\lambda\lambda}\rightarrow DN]}{\Gamma_{\text{Total}}}\sim11\%.
\end{eqnarray}
Therefore, $DN$ may be another decay channel for exploring the state $\Sigma_c|J^P=3/2^+,2\rangle_{\lambda\lambda}$.

The $\Sigma_c|J^P=5/2^+,2\rangle_{\lambda\lambda}$ may have a narrow width of
$\Gamma\sim25$ MeV. This state may have significant decay rates into $D^*N$, $\Sigma_c^*\pi$
and $\Sigma_c|J^P=5/2^-,2\rangle_{\lambda}\pi$ with comparable branching fractions $\sim 20-30\%$, according to the
predictions of the QPC model. Due to the narrow decay width, the state $\Sigma_c|J^P=5/2^+,2\rangle_{\lambda\lambda}$ has
a good potential to be observed in experiment.

For the last two $1D$-wave $\lambda$-mode states $\Sigma_c|J^P=5/2^+,3\rangle_{\lambda\lambda}$ and $\Sigma_c|J^P=7/2^+,3\rangle_{\lambda\lambda}$, they are probably narrow states with a total decay width of $\Gamma\sim(9-14)$ MeV predicted within the QPC model.
With the ChQM, their widths are predicted to be $\Gamma\sim30$ MeV, which is slightly broader than the prediction of the QPC model.
The predicted decay properties with the two strong decay models are very different from each other.
In the ChQM, both $\Sigma_c|J^P=5/2^+,3\rangle_{\lambda\lambda}$ and $\Sigma_c|J^P=7/2^+,3\rangle_{\lambda\lambda}$
dominantly decay into the $\Lambda_c\pi$ channel. However, in the QPC model,
the main decay channel of $\Sigma_c|J^P=5/2^+,3\rangle_{\lambda\lambda}$ is predicted to be $\Sigma_c|J^P=5/2^-,2\rangle_{\lambda}\pi$,
while the dominant decay channel of $\Sigma_c|J^P=7/2^+,3\rangle_{\lambda\lambda}$ is $\Sigma_c|J^P=3/2^-,1\rangle_{\lambda}\pi$.
If the predictions of the ChQM are reliable, both $\Sigma_c|J^P=5/2^+,3\rangle_{\lambda\lambda}$ and
$\Sigma_c|J^P=7/2^+,3\rangle_{\lambda\lambda}$ may be observed in the $\Lambda_c\pi$ channel.
If the predictions of the QPC model are reliable, they mainly decay into the $1P$-wave $\Sigma_c$
states via pionic decay processes.

\begin{figure*}[]
	\centering \epsfxsize=15 cm \epsfbox{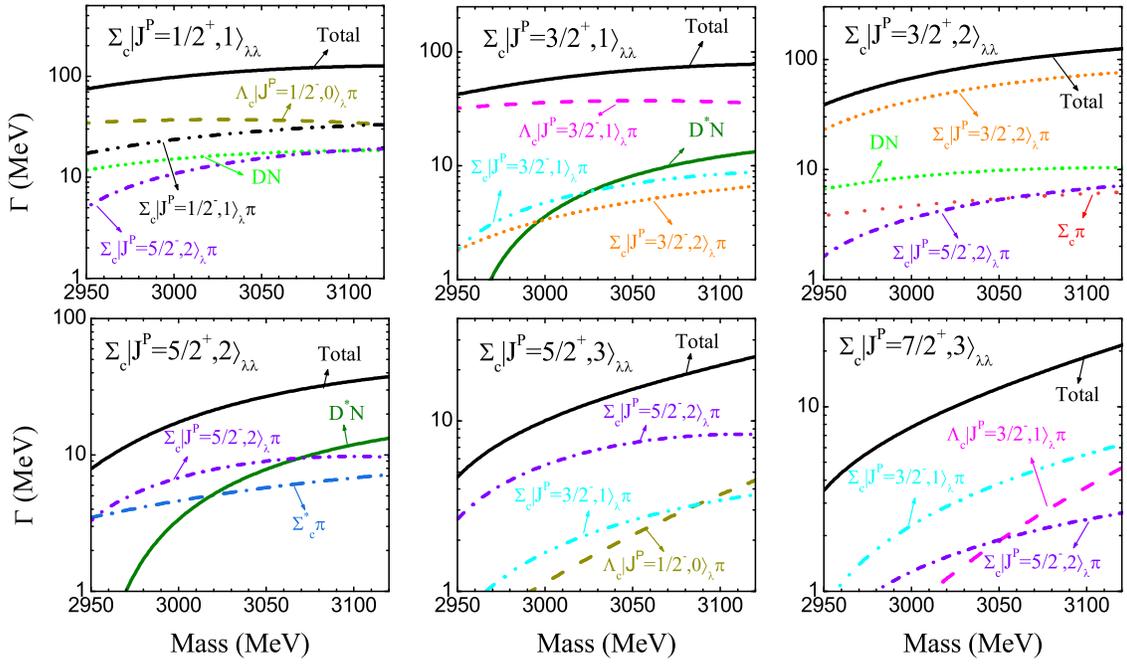}
	\caption{ Partial and total strong decay widths of the
		$1D$-wave $\lambda$-mode $\Sigma_c$ states as functions of their masses. Some decay channels are too small to show in figure }
\label{FIG4}
\end{figure*}

We also plot the partial decay widths of the $1D$-wave $\lambda$-mode $\Sigma_c$ baryons within the QPC model as a function of the mass in region of $M=(2950-3120)$ MeV. The sensitivities of the decay properties of these states to their masses are shown in Fig.~\ref{FIG4}. Within the masses varied in the region we considered in present work, the $1D$-wave $\lambda$-mode $\Sigma_c$ baryons have a large potential to be observed in their main decay channels in experiments.

\subsection{$1D$-wave $\rho$-mode excitations}

For the $1D$-wave $\rho$-mode $\Sigma_c$ baryons, there are also six states $\Sigma_c|J^P=1/2^+,1\rangle_{\rho\rho}$, $\Sigma_c|J^P=3/2^+,1\rangle_{\rho\rho}$, $\Sigma_c|J^P=3/2^+,2\rangle_{\rho\rho}$, $\Sigma_c|J^P=5/2^+,2\rangle_{\rho\rho}$, $\Sigma_c|J^P=5/2^+,3\rangle_{\rho\rho}$ and $\Sigma_c|J^P=7/2^+,3\rangle_{\rho\rho}$. So far, numerical values of their masses are still missing.
While, according to the quark model predictions~\cite{Bijker:2020tns,Yoshida:2015tia,Capstick:1986ter,Chen:2021eyk}
the mass of the $\rho$-mode excitations is about (70-150) MeV higher than that of the $\lambda$-mode excitations. Thus, we fix masses of the $1D$-wave $\rho$-mode states on the values which are 100 MeV higher than those of the corresponding $1D$-wave $\lambda$-mode states.
We predict the strong decay properties with both the QPC model and ChQM, and our results are collected in Table~\ref{Dlambda}.
It should be mentioned that the $DN$ and $D^*N$ decay channels are forbidden since
we adopt the simple harmonic oscillator wave functions in present work, although the masses of the $1D$-wave
$\rho$-mode $\Sigma_c$ states are above the threshold of $DN$ and $D^*N$.

\begin{figure*}[]
	\centering \epsfxsize=15 cm \epsfbox{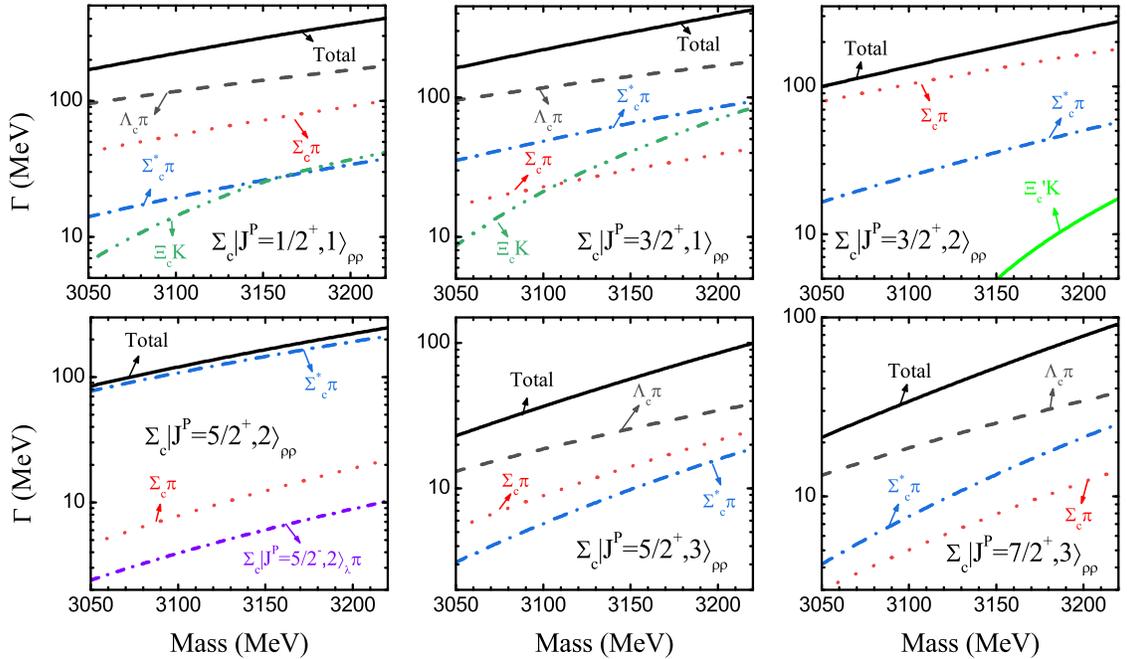}
	\caption{ Partial and total strong decay widths of the
		$1D$-wave $\rho$-mode $\Sigma_c$ states as functions of their masses. Some decay channels are too small to show in figure. }
\label{FIG5}
\end{figure*}

According to the QPC model predictions, it is obtained that the four states $\Sigma_c|J^P=1/2^+,1\rangle_{\rho\rho}$, $\Sigma_c|J^P=3/2^+,1\rangle_{\rho\rho}$, $\Sigma_c|J^P=3/2^+,2\rangle_{\rho\rho}$ and $\Sigma_c|J^P=5/2^+,2\rangle_{\rho\rho}$ may be broad states with a width of $\Gamma\sim(150-280)$ MeV; while the other two states $\Sigma_c|J^P=5/2^+,3\rangle_{\rho\rho}$ and $\Sigma_c|J^P=7/2^+,3\rangle_{\rho\rho}$ have a moderate width of around $\Gamma\sim(40-45)$ MeV, and may have good potential to be observed in forthcoming experiments. However, with the ChQM all of the six $\rho$-mode states are predicted to be
moderate states with a width of $\Gamma\sim(40-60)$ MeV.

For the four broad states predicted in the QPC model,
the $\Sigma_c|J^P=1/2^+,1\rangle_{\rho\rho}$ and $\Sigma_c|J^P=3/2^+,1\rangle_{\rho\rho}$ states
have similar decay properties, they mainly decay into $\Lambda_c\pi$, $\Sigma_c\pi$, $\Sigma_c^*\pi$, and $\Xi_cK$ channels;
while the other two states $\Sigma_c|J^P=3/2^+,2\rangle_{\rho\rho}$ and $\Sigma_c|J^P=5/2^+,2\rangle_{\rho\rho}$
have a similar decay width, they dominantly decay into the $\Sigma_c\pi$ and $\Sigma_c^*\pi$ channels, respectively.
Both $\Sigma_c|J^P=5/2^+,3\rangle_{\rho\rho}$ and $\Sigma_c|J^P=7/2^+,3\rangle_{\rho\rho}$
are most likely to be moderate states with a comparable width of $\Gamma\sim(40-50)$ MeV. They
have similar decay properties, and mainly decay into $\Lambda_c\pi$, $\Sigma_c\pi$
and $\Sigma_c^*\pi$ channels. To looking for the missing $1D$-wave $\rho$-mode $\Sigma_c$ baryons,
the $\Lambda_c\pi$, $\Sigma_c\pi$ and $\Sigma_c^*\pi$ channels are worth observing
around the mass range of $\sim3.1-3.2$ GeV in future experiments.

Considering the uncertainty of the masses of the $1D$-wave $\rho$-mode states,
we further investigate the strong decay widths of the $1D$-wave $\rho$-mode $\Sigma_c$ states with the QPC model as a function of the mass in Fig.~\ref{FIG5}. It is shown that the decay properties of the $1D$-wave $\rho$-mode  $\Sigma_c$ excitations are sensitive to their masses varying in region of $M=(3050-3220)$ MeV.

\section{Summary}

In this work, we systematically investigated the two-body strong decays of low-lying $1P$- and $1D$-wave
$\Sigma_c$ excitations in the QPC model within the $j$-$j$ coupling scheme. For a comparison, we also
give our predictions within the chiral quark model. The main theoretical results are summarized as follows.

In the $1P$-wave $\lambda$-mode states, the three states $\Sigma_c|J^{P}=1/2^-,0\rangle_{\lambda}$,
$\Sigma_c|J^{P}=3/2^-,2\rangle_{\lambda}$ and $\Sigma_c|J^{P}=5/2^-,2\rangle_{\lambda}$
dominantly decay into $\Lambda_c\pi$. The predicted decay widths
have a strong model dependency. In the QPC model, the $\Sigma_c|J^{P}=1/2^-,0\rangle_{\lambda}$
has a moderate width of $\Gamma\sim50$ MeV, while the other two states are very narrow states
with a width of $\Gamma\sim4-7$ MeV. However, in the ChQM, the $\Sigma_c|J^{P}=1/2^-,0\rangle_{\lambda}$
is predicted to be a narrow state, while both $\Sigma_c|J^{P}=3/2^-,2\rangle_{\lambda}$ and
$\Sigma_c|J^{P}=5/2^-,2\rangle_{\lambda}$ have a moderate width of $\Gamma\sim40$ MeV.
The $\Sigma_c(2800)$ structure may be contributed by the
$\Sigma_c|J^{P}=1/2^-,0\rangle_{\lambda}$ state or potentially arises from the
overlapping the three $1P$-wave $\Sigma_c$ states with different quantum numbers $J^P=1/2^-$,
$J^P=3/2^-$ and $J^P=5/2^-$. The $\Sigma_c|J^{P}=1/2^-,0\rangle_{\lambda}$ may a
large decay rate into the $DN$ channel. The observations of this channel may
be useful to distinguish the $\Sigma_c|J^{P}=1/2^-,0\rangle_{\lambda}$ from the other $P$-wave
states. The two $1P$-wave $\lambda$-mode states $\Sigma_c|J^{P}=1/2^-,1\rangle_{\lambda}$,
$\Sigma_c|J^{P}=3/2^-,1\rangle_{\lambda}$ dominantly decay into $\Sigma_c\pi$
and $\Sigma_c^*\pi$, respectively. Both the QPC model and ChQM give consistent
predictions of their decay properties. These two $1P$-wave states may be moderate
width states with a width of $\sim10$s MeV. They are most likely to be observed in
the $\Lambda_c\pi\pi$ invariant mass spectrum
around the energy range of $\sim2.75-2.85$ GeV in experiments.

The $1P$-wave $\rho$-mode states $\Sigma_c|J^{P}=1/2^-,1\rangle_{\rho}$ and
$\Sigma_c|J^{P}=3/2^-,2\rangle_{\rho}$ dominantly decay into $\Sigma_c\pi$
and $\Sigma_c^*\pi$, respectively. They may have a broad width of $\Gamma\sim(100-200)$ MeV.
To looking for these $1P$-wave $\rho$-mode states, the $\Lambda_c\pi\pi$ invariant mass spectrum
around the energy range of $\sim2.85-2.95$ GeV is worth to observing in future experiments

For the $1D$-wave $\lambda$-mode $\Sigma_c$ baryons, the main decay channels predicted
within the QPC model and ChQM are roughly consistent with each other. Most of these
$1D$-wave $\lambda$-mode states have a relatively narrow width of $\sim10$s MeV, and
dominantly decay into the $1P$-wave $\Sigma_c$ and/or $\Lambda_c$ states.
According to the QPC model predictions, both $\Sigma_c|J^{P}=1/2^+,1\rangle_{\lambda\lambda}$
and $\Sigma_c|J^{P}=3/2^+,2\rangle_{\lambda\lambda}$ have sizable decay rates
decaying into the $DN$ channel, which may be an interesting decay channel for
experimental observations. In the ChQM, both $\Sigma_c|J^{P}=5/2^+,3\rangle_{\lambda\lambda}$
and $\Sigma_c|J^{P}=7/2^+,3\rangle_{\lambda\lambda}$ mainly decay into the
$\Lambda_c\pi$ channel, which is worth to observing in future experiments.

For the $1D$-wave $\rho$-mode $\Sigma_c$ baryons, both $\Sigma_c|J^{P}=1/2^+,1\rangle_{\rho\rho}$ and
$\Sigma_c|J^{P}=3/2^+,1\rangle_{\rho\rho}$ have similar decay properties.
They have large decay rates into the $\Lambda_c\pi$,
$\Sigma_c\pi$, $\Sigma_c^*\pi$ and $\Xi_cK$ channels.
According to the QPC model predictions, they may be broad states with a width of about $\Gamma\sim(150-280)$ MeV,
which is about a factor of $4-5$ larger than that predicted with ChQM. The $\Sigma_c|J^P=3/2^+,2\rangle_{\rho\rho}$ and $\Sigma_c|J^P=5/2^+,2\rangle_{\rho\rho}$ mainly decay into the $\Sigma_c\pi$ and $\Sigma_c^*\pi$ channels,
respectively. They have a similar decay width of $\Gamma\sim(50-180)$ MeV, which shows some model dependencies.
The $\Sigma_c|J^P=5/2^+,3\rangle_{\rho\rho}$ and $\Sigma_c|J^P=7/2^+,3\rangle_{\rho\rho}$
are most likely to be moderate states with a comparable width of $\Gamma\sim(40-50)$ MeV. They
have similar decay properties, and mainly decay into $\Lambda_c\pi$, $\Sigma_c\pi$
and $\Sigma_c^*\pi$ channels. To looking for the missing $1D$-wave $\rho$-mode $\Sigma_c$ baryons,
the $\Lambda_c\pi$, $\Sigma_c\pi$ and $\Sigma_c^*\pi$ channels are worth observing
around the mass range of $\sim3.1-3.2$ GeV in future experiments.

\section*{Acknowledgements }

We would like to thank Kai-Lei Wang for very helpful discussions. This work is supported by the National Natural Science Foundation of China under Grants No.12005013, No.12175065, No.12235018 and No.11947048.

\bibliography{Sigmacv6}

	\end{document}